\documentclass[journal]{IEEEtran}
\usepackage{subfigure}
\usepackage{cite}
\usepackage{citesort}
\usepackage{amsmath,epsfig,bm}
\usepackage{pifont}
\usepackage{paralist}
\usepackage{graphicx}
\usepackage{amssymb}
\usepackage{amsmath}
\usepackage{cancel}
\usepackage[normalem]{ulem}
\usepackage[dvips]{color}
\usepackage{algorithm}

\DeclareMathOperator*{\argmin}{argmin}

\usepackage{pifont}
\usepackage{graphicx}
\usepackage{amssymb}
\usepackage{amsmath}
\usepackage{cancel}
\usepackage[normalem]{ulem}
\usepackage[dvips]{color}
\usepackage{algorithm}
\usepackage{algorithmic}

\newtheorem{theorem}{\noindent \textbf{Theorem}}

\newtheorem{corollary}{\noindent \textbf{Corollary}}
\begin{document}
%%%%%%%%%%%%%%%%%%%%%%%%%%%%%%%%%%%%%%%%%%%%%%%%%
\title{Secrecy Performance Analysis of Location-Based Beamforming in Rician Wiretap Channels}
%%%%%%%%%%%%%%%%%%%%%%%
\author{Shihao~Yan and Robert~Malaney

\thanks{S. Yan and R. Malaney are with the School of Electrical Engineering and Telecommunications, The University of New South Wales, Sydney, NSW 2052, Australia (Emails: shihao.yan@unsw.edu.au; r.malaney@unsw.edu.au).}
\thanks{This work was funded by The University of New South Wales and Australian Research Council Grant DP120102607.}}

%%%%%%%%%%%%%%%%%%%%%%%
\maketitle

%%%%%%%%%%%%%%%%%%%%%%%%%%%%%%%%%%%%%%%%%%%%%%%%%%%
% ABSTRACT
%%%%%%%%%%%%%%%%%%%%%%%%%%%%%%%%%%%%%%%%%%%%%%%%%%%
\begin{abstract}
We propose a new optimal Location-Based Beamforming (LBB) scheme for the  wiretap channel, where both the main channel and the eavesdropper's channel are subject to Rician fading. In our LBB scheme the two key inputs are the location of the legitimate receiver and the location of the potential eavesdropper. Notably, our  scheme does not require as direct inputs any channel state information of the main channel or the eavesdropper's channel, making it easy to deploy in a host of  application settings in which the location inputs are known. Our beamforming solution  assumes a multiple-antenna transmitter, a multiple-antenna eavesdropper, and a single-antenna receiver, and its aim is to maximize  the physical layer security of the channel. To obtain our solution we first derive the secrecy outage probability  of the LBB scheme in a closed-form expression that is valid for arbitrary values of the Rician $K$-factors of the main channel and the eavesdropper's channel. Using this expression we then determine the location-based  beamformer solution that minimizes the secrecy outage probability. To assess the usefulness of our new scheme, and to quantify the value of the location information to the beamformer, we compare our scheme to other schemes, some of which do not utilize any location information. Our new  beamformer solution provides  optimal physical layer security for a wide range of location-based applications.
\end{abstract}


\section{Introduction}

%Research into wireless physical-layer-based secure communications has recently gained considerable interest.
Physical layer security guarantees secrecy regardless
of an eavesdropper's computational capability and does not require complex key distribution and management \cite{shiu2011physical}.
In early  studies \cite{shannon1949communication,wyner1975the}, a
wiretap channel model was proposed as the fundamental system model to
examine such physical layer security in single-input single-output systems.
In the wiretap channel, an
eavesdropper (Eve) overhears the wireless
communication between a transmitter (Alice) and an intended
receiver (Bob).
More recently, motivated by multiple-input multiple-output (MIMO) techniques, physical layer security in MIMO wiretap channels has garnered much interest (e.g., \cite{khisti2010secure,mukherjee2011robust,goel2008guaranteeing,zhou2010secure,yang2013transmit,yan2014transmit}).
However, many of the works in MIMO-based physical layer security assume the (instantaneous) CSI of the \emph{main channel} (the channel between Alice and Bob) is perfectly known by Alice or Bob (e.g., \cite{khisti2010secure,goel2008guaranteeing,zhou2010secure}).
This assumption is usually very difficult to justify in practice (e.g., in  massive MIMO techniques the CSI of a channel cannot be perfectly known even to a receiver due to pilot contamination issues \cite{rusek2013scaling,larsson2014massive,marzetta2010non,yin2013acoordinated}).
Another assumption adopted in the literature is that the CSI of the \emph{eavesdropper's channel} (the channel between Alice and Eve) is known to Alice, which is even harder to justify in practice.

However, there are many circumstances where \emph{location information} of Bob and Eve could be available. For example, in some specific military application scenarios, Alice may obtain Bob's location through direct communications, and Eve's location through some (possibly\emph{ a priori}) surveillance. Other circumstances could be where Bob and Eve are  known users of the system (but still requiring  secret communications on an individual basis), and their location information is routinely broadcasted as per the requirements of the network protocol. Examples of such circumstances would be in IEEE 1609.2 for vehicular networks, or in some location-based social-media applications.

Regardless of the application scenario, the main point we focus on here is that if there is a line-of-sight (LOS) component in the main channel or the eavesdropper's channel,  it is possible to utilize location information directly in order to enhance the physical layer security. More specifically, we propose and analyze a new Location-Based Beamforming (LBB) scheme in the wiretap channel, where both the main channel and the eavesdropper's channel are subject to Rician fading. Our scheme does not require the CSI of either the main channel or the eavesdropper's channel - thus making it quite general, as well as pragmatic. The basic\emph{ modus operandi} of the scheme we propose is that given the input locations of Bob and Eve, we output the optimal beamformer  solution and  the security level (the secrecy outage probability) associated with this solution.\footnote{Although our scheme works for any input locations. It is possible that the secrecy outage probability approaches one (e.g., as Bob moves further from Alice whilst Eve moves closer). We leave it to the system operator to decide whether the secrecy outage predicted justifies the sending of data.} Detailing how these outputs are determined  forms the core of our work.

Surprisingly, there has been little previous work in this area, with the closest works perhaps those of  \cite{li2011ergodic} and \cite{ferdinand2013phisical}.
In \cite{li2011ergodic}, the ergodic secrecy rate was examined for multiple-antenna wiretap channels with Rician fading. However, in \cite{li2011ergodic} it was assumed that the CSI of the main channel was perfectly known by Alice.
%to determine the optimal input covariance matrix that maximizes the ergodic secrecy rate.
The work of \cite{ferdinand2013phisical} analyzed the secrecy performance of orthogonal space-time block codes when the main channel is assumed to be subject to Rician fading. But the eavesdropper's channel was assumed to be subject to Rayleigh fading in \cite{ferdinand2013phisical} and therefore Eve's location information was not that useful.

The direction of this paper and our  contributions are summarized as follows. (i) We first derive the secrecy outage probability  of the LBB scheme in a closed-form expression, which is valid for arbitrary values of the Rician $K$-factors of the main channel and the eavesdropper's channel. (ii) We then determine the optimal location-based beamformer and the minimum secrecy outage probability for the scheme.
(iii) In order to fully appreciate the gains of the LBB scheme, we also analyze, for comparison, the secrecy performance of a Non-Beamforming (NB) scheme.
(iv) As a final comparison, we also consider the effect on the LBB scheme of the full CSI of Bob being made available to Alice, and the effect of Eve's location information becoming untrustworthy.

The rest of this paper is organized as follows. Section \ref{System Model} details our system model; Section~\ref{sec_LBB} provides our analytical solutions; Section~\ref{sec_numerical} provides numerical simulations; and  Section \ref{sec_conclusion} draws concluding remarks. Secrecy performances of the comparison schemes  are provided in Appendices. We adopt the following notations in this work. Scalar variables are denoted by italic symbols. Vectors and matrices are denoted by lower-case and upper-case boldface symbols, respectively. Given a complex number $z$, $|z|$ denotes the modulus of $z$. Given a complex vector $\mathbf{x}$,
$\|\mathbf{x}\|$ denotes the Euclidean norm, $\mathbf{x}^{T}$ denotes the transpose of $\mathbf{x}$, $\mathbf{x}^{\dag}$ denotes the conjugate transpose of $\mathbf{x}$, and $\text{Re}(\mathbf{x})$ denotes the real part of $\mathbf{x}$. The $L\times L$ identity matrix
is referred to as $\mathbf{I}_{L}$ and $\mathbb{E}[\cdot]$ denotes expectation.

\section{System Model}\label{System Model}

\begin{figure}[!t]
    \begin{center}
   {\includegraphics[width=3.2in]{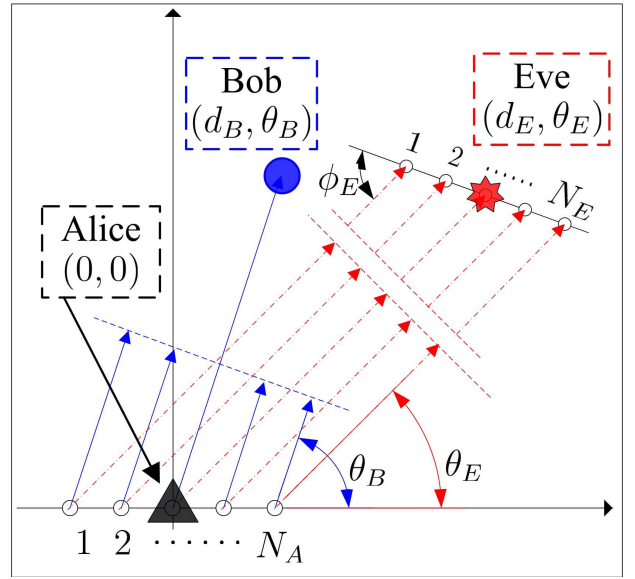}}
    \caption{Illustration of the Rician wiretap channel of interest.}\label{fig:system}
    \end{center}
\end{figure}

Our LBB scheme was examined for the simpler case of a pure LOS channel in one of our previous works \cite{yan2014line}. Here, we expand on that simple scenario by considering more generic and realistic channel conditions. That is,
 we will assume $K_B > 0$ and $K_E > 0$, where $K_B$ and $K_E$ are the Rician $K$-factors of the main channel and the eavesdropper's channel, respectively.
The wiretap channel of interest is illustrated in Fig.~\ref{fig:system}, where  Alice and Eve are equipped with uniform linear arrays (ULAs) with $N_A$ and $N_E$ antenna elements,\footnote{We will assume $N_E$ is also known to Alice. This is reasonable in circumstances where Alice can determine physical constraints on the size of an eavesdropper's antenna, knowledge of which, coupled to the known frequency of transmission, can allow for a reliable upper bound on $N_E$ to be set. If an upper bound on $N_E$ is set, then our solutions become bounds (worst case scenarios). In other circumstances, where Eve is at times a legitimate user, we can assume $N_E$ is known.} respectively; and Bob is equipped with a single antenna. As we will show later, our analysis provided in this work is also valid for other antenna arrays beyond ULAs at Eve. We assume that Alice, Bob, and Eve are static.

As shown in Fig.~\ref{fig:system}, we adopt the polar coordinate system, where Alice's location is selected as the origin, Bob's location is denoted as $(d_B, \theta_B)$, and Alice's location is denoted as $(d_E, \theta_E)$. For presentation convenience, without other statements we assume that the coordinate system is set up such that $0 \leq \theta_B \leq \pi$ and $0 \leq \theta_E \leq \pi$.
The orientation of the ULA at Alice is also shown in this figure. We also assume that
the main channel and the eavesdropper's channel are subject to quasi-static Rician fading with equal block length but different Rician $K$-factors, and that a $K$-factor map ($K$ as a function of locations) is known in the vicinity of Alice via some \emph{a priori} measurement campaigns.
We further assume that the CSI of the main channel is unknown to Alice, but that Bob's location is known to Alice.\footnote{We note that using Bob's location  saves  feedback overhead relative to use of the CSI of the main channel. This is due to the following two facts: (i) the CSI varies during different fading blocks and has to be fed back for each fading block, meanwhile the location information only has to be fed back once for a static Bob; and (ii) the CSI is an $N_A$-dimension complex vector ($2N_A$ variables embedded), meanwhile Bob's location is determined by only two real numbers.} Additional assumptions are that Eve knows the CSI of the eavesdropper's channel and the beamformer adopted by Alice;  that Eve applies Maximum Ratio Combining (MRC) in order to maximize the probability of successful eavesdropping \cite{yang2013transmit,yan2014transmit}; and that Eve's location is known to Alice.
%This last assumption can be justified by the fact that Alice may know Eve's location through some surveillance program (e.g., an enemy position previously identified).
As we discuss later, our analysis also covers the case where Eve's location is unavailable at Alice.

As per the aforementioned assumptions, the $1 \times N_A$ main channel vector is given by
\begin{align}\label{h_definition}
\mathbf{h} = \sqrt{\frac{K_B}{1+K_B}} \mathbf{h}_o + \sqrt{\frac{1}{1+K_B}} \mathbf{h}_r,
\end{align}
where $\mathbf{h}_o$ is the LOS component, and $\mathbf{h}_r$ is the scattered component. The entries of $\mathbf{h}_r$ are independent and identically distributed (i.i.d) circularly-symmetric complex Gaussian random variables with zero mean and unit variance, i.e., $\mathbf{h}_r \sim \mathcal{CN}(0, \mathbf{I}_{N_A})$. Denoting $\rho_A$ as the space between two antenna elements of the ULA at Alice, $\mathbf{h}_o$ is given by \cite{tsai2004ber}
\begin{align}\label{h_o_definition}
\bm{h}_{o} = \left[1,\cdots,\exp(j (N_A -1)\tau_A \cos \theta_B)\right],
\end{align}
where $\tau_A = 2 \pi f_0 \rho_A /c$, $f_0$ is the carrier frequency, and $c$ is the speed of propagation of the plane wave.
The $N_E \times N_A$ eavesdropper's channel matrix is given by
\begin{align}\label{g_definition}
\mathbf{G} = \sqrt{\frac{K_E}{1+K_E}} \mathbf{G}_o + \sqrt{\frac{1}{1+K_E}} \mathbf{G}_r,
\end{align}
where $\mathbf{G}_o$ is the LOS component, and $\mathbf{G}_r$ is the scattered component represented by a matrix with i.i.d circularly-symmetric complex Gaussian random variables with zero mean and unit variance. Given the locations of Alice and Eve, $\mathbf{G}_o$ can be written as \cite{taricco2011on}
\begin{align}\label{G_o_definition}
\mathbf{G}_o = \mathbf{r}_o^T \mathbf{g}_o
\end{align}
where $\mathbf{r}_o$ and $\mathbf{g}_o$ are the  array responses at Eve and Alice, respectively, which are given by
\begin{align}
\mathbf{r}_{o} &= \left[1,\cdots,\exp(-j (N_E -1)\tau_E \cos \phi_E)\right], \label{r_o} \\
\mathbf{g}_{o} &= \left[1,\cdots,\exp(j (N_A -1)\tau_A \cos \theta_E)\right].\label{g_o_definition}
\end{align}
In \eqref{r_o}, we have $\tau_E = 2 \pi f_0 \rho_E /c$, where $\rho_E$ is the space between two antenna elements of the ULA at Eve, and $\phi_E$ is the direction of arrival from Eve to Alice which is dependent on the orientation of the ULA at Eve. As we show later, the signal-to-noise ratio (SNR) of the eavesdropper's channel is independent of $\phi_E$ when Eve utilizes MRC to combine the received signals. As such, the secrecy performance of the LBB scheme does not depend on $\phi_E$ and thus Alice does not have to know $\phi_E$.

The received signal at Bob is given by
\begin{align}\label{y_received}
y = \sqrt{{g(d_B)}} \mathbf{h}\mathbf{b} x + n_B,
\end{align}
where $g(d_B)$ is the path loss component of the main channel given by $g(d_B) = \left(c/4\pi f_0 d_0\right)^2({d_0}/{d_B})^{\eta_B}$ ($d_0$ is a reference distance and $\eta_B$ is the path loss exponent\footnote{The path loss exponent $\eta_B$ is dependent on the Rician $K$-factor $K_B$. For example, $\eta_B \rightarrow 2$ as $K_B \rightarrow \infty$. For simplicity, we assume $\eta_B$ is known to Alice since $K_B$ is known. This declaration also applies to the path loss exponent of the eavesdropper's channel $\eta_E$ and the Rician $K$-factor $K_E$.} of the main channel), $\mathbf{b}$ is a normalized beamformer (i.e., $\|\mathbf{b}\| = 1$), $x$ is the Gaussian distributed information bearing signal satisfying $\mathbb{E}[|x|^2] = P$ ($P$ is the total transmit power of Alice\footnote{It is straightforward to prove that the secrecy outage probability is a monotonically decreasing function of Alice's transmit power for given locations of Bob and Eve. As such, we assume that Alice always sets her transmit power at the maximum value $P$.}), and $n_B$ is the additive white Gaussian noise of the main channel with zero mean and variance $\sigma_B^2$.
Likewise, the received signal at Eve is given by
\begin{align}\label{z_received}
\mathbf{z} = \sqrt{{g(d_E)}} \mathbf{G}\mathbf{b} x + \mathbf{n}_E,
\end{align}
where ${{g(d_E)}}$ is the path loss component of the eavesdropper's channel given by $g(d_E) = \left(c/4\pi f_0 d_0\right)^2({d_0}/{d_E})^{\eta_E}$ ($\eta_E$ is the path loss exponent of the eavesdropper's channel), and $\mathbf{n}_E$ is the additive white Gaussian noise vector of the eavesdropper's channel with zero mean and variance matrix  $\sigma_E^2 \mathbf{I}_{N_E}$, i.e., $\mathbf{n}_E \sim \mathcal{CN}(\mathbf{0}, \sigma_E^2\mathbf{I}_{N_E})$

Then, the SNR of the main channel is given by
\begin{align}\label{SNR_B}
\gamma_B = \frac{P g(d_B) |\mathbf{h}\mathbf{b}|^2}{\sigma_B^2} = \overline{\gamma}_B |\mathbf{h}\mathbf{b}|^2,
\end{align}
where $\overline{\gamma}_B$ is defined as $\overline{\gamma}_B \triangleq {P g(d_B)}/{\sigma_B^2}$.
Assuming Eve applies MRC to combine the received signals at different antennas, the SNR of the eavesdropper's channel is given by
\begin{align}\label{SNR_E}
\gamma_E = \frac{P g(d_E) \|\mathbf{G}\mathbf{b}\|^2}{\sigma_E^2} = \overline{\gamma}_E \|\mathbf{G}\mathbf{b}\|^2,
\end{align}
where $\overline{\gamma}_E$ is defined as $\overline{\gamma}_E \triangleq {P g(d_E)}/{\sigma_E^2}$.

\section{Location-based Beamforming Scheme}\label{sec_LBB}

In this section, we first examine the secrecy performance of our proposed LBB scheme in terms of the secrecy outage probability and the probability of non-zero secrecy capacity. We then determine the optimal  location-based beamformer of the LBB scheme that minimizes the secrecy outage probability.

\subsection{Preliminaries}

In order to derive the secrecy performance metrics of our  scheme (e.g., the secrecy outage probability), we first derive the probability density functions (pdfs) of $\gamma_B$ and $\gamma_E$. Without loss of generality, we  derive such pdfs for a general $\mathbf{b}$, which is independent of $\mathbf{h}_r$ and $\mathbf{G}_r$. To this end, we first determine the distribution type of $|\mathbf{h}\mathbf{b}|$. As per \eqref{h_definition}, we have
\begin{align}\label{h_tilde}
\mathbf{h} \mathbf{b}= \underbrace{\sqrt{\frac{K_B}{1+K_B}} \mathbf{h}_o \mathbf{b}}_{\tilde{h}_o} + \underbrace{\sqrt{\frac{1}{1+K_B}} \mathbf{h}_r \mathbf{b}}_{\tilde{h}_r}.
\end{align}
Since $\mathbf{b}$ is independent of $\mathbf{h}_r$, $\tilde{h}_r$ is still a circularly-symmetric complex Gaussian random variable. Noting that $\tilde{h}_o$ is deterministic, we conclude that $|\mathbf{h}\mathbf{b}|$ follows a Rician distribution. We next determine the parameters of this Rician distribution. Following \eqref{h_tilde}, we have
\begin{align}\label{tilde_1}
|\tilde{h}_o|^2 = \frac{K_B}{1+K_B}|\mathbf{h}_o \mathbf{b}|^2
\end{align}
and
\begin{align}\label{tilde_2}
\mathbb{E}[|\tilde{h}_r|^2] = \frac{1}{1+K_B}\mathbb{E}[|\mathbf{h}_r \mathbf{b}|^2] = \frac{1}{1+K_B}.
\end{align}
We note that $|\tilde{h}_o|^2$ is the power of the LOS (deterministic) component and $\mathbb{E}[|\tilde{h}_r|^2]$ is the average power of the non-LOS (random) component. As such, we conclude that $|\mathbf{h}\mathbf{b}|$ follows a Rician distribution with $\widetilde{K}_B$ and $\widetilde{\overline{\gamma}}_B$ as the Rician $K$-factor and total power, respectively, where
$\widetilde{K}_B$ and $\widetilde{\overline{\gamma}}_B$ are given by
\begin{align}
\widetilde{K}_B &\triangleq \frac{|\tilde{h}_o|^2}{\mathbb{E}[|\tilde{h}_r|^2]} = |\mathbf{h}_o \mathbf{b}|^2 K_B,\label{widetilde_KB}\\
\widetilde{\overline{\gamma}}_B &\triangleq \mathbb{E}[\gamma_B] \!=\! \overline{\gamma}_B \left(|\tilde{h}_o|^2 + \mathbb{E}[|\tilde{h}_r|^2]\right) \!=\! \frac{\left(K_B |\mathbf{h}_o \mathbf{b}|^2 + 1\right)\overline{\gamma}_B} {1+K_B}.
\end{align}
The pdf of Rician random variables involves the zero-order modified Bessel function of the first kind, which is not suitable for further analysis (e.g., deriving the secrecy outage probability). To make progress, it is convenient to interpret the Rician fading as a special case of  Nakagami fading. As such, the pdf of $\gamma_B$ is approximated as \cite{goldsmith2005wireless}
\begin{align}\label{pdf_SNR_B}
f_{\gamma_B}(\gamma) = \left(\frac{\widetilde{m}_B}{\widetilde{\overline{\gamma}}_B}\right)^{\widetilde{m}_B} \frac{\gamma^{\widetilde{m}_B-1}}{\Gamma(\widetilde{m}_B)}\exp\left(\frac{-\widetilde{m}_B \gamma}{\widetilde{\overline{\gamma}}_B}\right),
\end{align}
where $\widetilde{m}_B$ is the Nakagami fading parameter given by $\widetilde{m}_B = (\widetilde{K}_B +1)^2/(2 \widetilde{K}_B +1)$ and $\Gamma(\mu) = \int_0^{\infty}e^{-t} t^{\mu-1} dt$, $\text{Re}(\mu) >0$, is the Gamma function.

Following \eqref{SNR_E}, the SNR of the eavesdropper's channel can be rewritten as
\begin{align}
\gamma_E = \sum_{i=1}^{N_E} \gamma_{E,i},
\end{align}
where $\gamma_{E,i} = \overline{\gamma}_E |\mathbf{g}_i\mathbf{b}|^2$, $\mathbf{g}_i$ is the $1 \times N_A$ channel vector between Eve's $i$-th antenna and Alice, i.e., $\mathbf{g}_i$ is the $i$-th row of $\mathbf{G}$. As per \eqref{g_definition}, we have
\begin{align}
\mathbf{g}_i = \sqrt{\frac{K_E}{1+K_E}} \epsilon_i \mathbf{g}_o + \sqrt{\frac{1}{1+K_E}} \mathbf{g}_{r,i},
\end{align}
where $\epsilon_i = e^{-j(i-1)\tau_E \cos \phi_E}$ and $\mathbf{g}_{r,i}$ is the $i$-th row of $\mathbf{G}_r$. For any value of $i$ ($i = 1, 2, \dots, N_E$), we have
\begin{align}\label{SNR_E_fact}
|\epsilon_i \mathbf{g}_o \mathbf{b}| = |\mathbf{g}_o \mathbf{b}|.
\end{align}
As such, following a  procedure similar to that used in  obtaining $f_{\gamma_B}(\gamma)$, the pdf of $\gamma_{E,i}$ can be approximated as
\begin{align}\label{pdf_SNR_E_i}
f_{\gamma_{E,i}}(\gamma) = \left(\frac{\widetilde{m}_E}{\widetilde{\overline{\gamma}}_E}\right)^{\widetilde{m}_E} \frac{\gamma^{\widetilde{m}_E-1}}{\Gamma(\widetilde{m}_E)}\exp\left(\frac{-\widetilde{m}_E \gamma}{\widetilde{\overline{\gamma}}_E}\right),
\end{align}
where $\widetilde{m}_E$ is given by $\widetilde{m}_E = (\widetilde{K}_E +1)^2/(2 \widetilde{K}_E +1)$, $\widetilde{K}_E$ is given by $\widetilde{K}_E = |\mathbf{g}_o\mathbf{b}|^2 K_E$, and $\widetilde{\overline{\gamma}}_E$ is given by
\begin{align}\label{average_SNR_E}
\widetilde{\overline{\gamma}}_E \triangleq \mathbb{E}[\gamma_E] = \frac{\left(K_E |\mathbf{g}_o \mathbf{b}|^2 + 1\right)\overline{\gamma}_E} {1+K_E}.
\end{align}
Since the $\gamma_{E,i}$ are independent, following \eqref{average_SNR_E} the pdf of $\gamma_E$ can be approximated as
\begin{align}\label{pdf_SNR_E}
f_{\gamma_E}(\gamma) = \left(\frac{\widetilde{m}_E}{\widetilde{\overline{\gamma}}_E}\right)^{N_E \widetilde{m}_E} \frac{\gamma^{N_E \widetilde{m}_E-1}}{\Gamma(N_E \widetilde{m}_E)}\exp\left(\frac{-\widetilde{m}_E \gamma}{\widetilde{\overline{\gamma}}_E}\right).
\end{align}

Following \eqref{SNR_E_fact}, we note that $\gamma_E$ is independent of $\mathbf{r}_o$. This  indicates that the SNR at Eve is independent of $\phi_E$ when Eve adopts MRC to combine received signals (we do not need to know the orientation of the ULA at Eve for our analysis). This also reveals that the SNR at Eve is independent of the type of antenna array at Eve (e.g., other antenna arrays beyond ULAs) since different antenna arrays only impact $\mathbf{r}_o$. As such, our following analysis is also valid for other antenna arrays at Eve (e.g., non-uniform linear arrays, circular arrays, rectangle arrays).

\subsection{Secrecy Performance of the LBB Scheme}

In the wiretap channel, the secrecy capacity is defined as
\begin{equation}\label{secrecy_capacity}
C_{s}=\left\{
\begin{array}{ll}
C_{B}-C_{E}~\;, &  \mbox{$\gamma_{B}>\gamma_{E}$}\\
0~\;, & \mbox{$\gamma_{B}\leq\gamma_{E}$},\;
\end{array}
\right.
\end{equation}
where $C_{B}=\log_{2}\left(1+\gamma_{B}\right)$ is the capacity of the main channel and $C_{E}=\log_{2}\left(1+\gamma_{E}\right)$ is the capacity of the eavesdropper's channel. Since $C_B$ and $C_E$ are unavailable at Alice, the perfect secrecy cannot be guaranteed in the wiretap channel of interest. For this reason we adopt the secrecy outage probability  and the probability of non-zero secrecy capacity as our secrecy performance metrics. The secrecy outage probability is defined as the probability of the secrecy capacity $C_s$ being less than the target secrecy
 rate $R_s$ (bits/channel-use), which can be formulated as \cite{yang2013transmit,yan2014transmit}\footnote{The secrecy outage probability is the most common metric used in physical layer security when  CSI on the channels is unavailable at Alice. However, it is important to note this metric does not distinguish between reliability and security \cite{zhou2011rethinking}.}
\begin{align}\label{outage_probability}
&\mathcal{O}\left(R_{s}\right)=\Pr\left(C_{s}<R_{s}\right)\notag\\
&=\int_0^{\infty}f_{\gamma_E}(\gamma_E)\left[\int_{0}^{2^{R_s}(1+\gamma_E) -1}f_{\gamma_B}(\gamma_B)d\gamma_B\right] d\gamma_E.
\end{align}

With regard to the secrecy performance of the LBB scheme, we first provide the following theorem.

\begin{theorem}\label{theorem1}
The secrecy outage probability of the LBB scheme for a given $R_s$ is
\begin{align}\label{secercy_outage_final}
&\mathcal{O} (R_s) =\notag\\
&\frac{ \widetilde{m}_B^{\widetilde{m}_B}\widetilde{m}_E^{N_E \widetilde{m}_E}2^{\widetilde{m}_B R_s}}{\Gamma(N_E \widetilde{m}_E)\widetilde{\overline{\gamma}}_B^{-N_E\widetilde{m}_E}
\widetilde{\overline{\gamma}}_E^{-\widetilde{m}_B}}
%------
\sum_{n = 0}^{+\infty}\frac{2^{n R_s}\exp\left(\!-\!\frac{\widetilde{m}_B \left(2^{R_s}\!-\!1\right)}{\widetilde{\overline{\gamma}}_B}\right)}
{\widetilde{m}_B^{\!-\!n}\widetilde{\overline{\gamma}}_B^n \Gamma(\widetilde{m}_B +n+1)} \times \notag\\
%===================================================================
&\sum_{l = 0}^{+\infty}\frac{\binom{\widetilde{m}_B\!+\!n}{l}\left({2^{R_s}\!\!-\!\!1}\right)^l
\left(\widetilde{\overline{\gamma}}_B \widetilde{\overline{\gamma}}_E\right)^{n\!-\!l}\Gamma_G(\widetilde{m}_B \!+\! N_E\widetilde{m}_E \!+\!n\!-\!l)}
{\!{2^{l R_s}}\!\left(2^{R_s}\widetilde{m}_B \widetilde{\overline{\gamma}}_E + \widetilde{m}_E \widetilde{\overline{\gamma}}_B \right)^{\widetilde{m}_B \!+\! N_E\widetilde{m}_E \!+\!n\!-\!l}},
\end{align}
where $\Gamma_G(\cdot)$ is the generalized gamma function (also valid for negative integers), which is given by \cite{fisher2012some}
\begin{align}
\Gamma_G(\alpha)\!=\!\left\{
\begin{array}{ll}
\frac{(-1)^{\!-\!\alpha}}{(\!-\!\alpha)!}
\left(\sum_{i=1}^{\!-\!\alpha}\frac{1}{i}\!+\!\alpha\right), &  \mbox{$\alpha$ is a negative integer},\\
\Gamma(\alpha), & \mbox{otherwise}.
\end{array}
\right.
\end{align}
\end{theorem}
\begin{IEEEproof}
Substituting \eqref{pdf_SNR_B} into \eqref{outage_probability}, $\mathcal{O}(R_s)$ is derived as
\begin{align}\label{outage_probability_1}
\mathcal{O}(R_s)= \int_0^{\infty}f_{\gamma_E}(\gamma_E) \frac{\gamma\left(\widetilde{m}_B, \frac{2^{R_s}(1+\gamma_E) -1}{\widetilde{m}_B^{-1}\widetilde{\overline{\gamma}}_B}\right)}{\Gamma(\widetilde{m}_B)} d\gamma_E,
\end{align}
where $\gamma\left(\alpha, \mu\right) = \int_0^{\mu} e^{-t} t^{\alpha-1} d t$, $\text{Re}\{\alpha\} > 0$, is the lower incomplete gamma function. In order to obtain the result in \eqref{outage_probability_1}, we have utilized the following identity \cite[Eq. (3.381.1)]{gradshteuin2007table}
\begin{align}\label{integral_1}
\int_0^{u} t^{\nu -1} e^{-\mu t} d t = \mu^{-\nu} \gamma(\nu, \mu u).
\end{align}
To make progress, we adopt the following identity to expand $\gamma\left(\alpha, \mu\right)$ \cite[Eq. (8.354.1)]{gradshteuin2007table}
\begin{align}\label{gamma_Expand}
\gamma\left(\alpha, \mu\right) = \sum_{n = 0}^{+\infty}\frac{\Gamma(\alpha) \mu^{\alpha+n}e^{-\mu}}{\Gamma(\alpha + n +1)}.
\end{align}
As per \eqref{gamma_Expand}, we have
\begin{align}\label{gamma_Expand_Exact}
&\gamma\left(\widetilde{m}_B, \frac{2^{R_s}(1+\gamma_E) -1}{\widetilde{m}_B^{-1}\widetilde{\overline{\gamma}}_B}\right) \notag \\
%==================================================================-
&= \sum_{n = 0}^{+\infty}\frac{\Gamma(\widetilde{m}_B)\left(\frac{2^{R_s}(1+\gamma_E) -1}{\widetilde{m}_B ^{-1}\widetilde{\overline{\gamma}}_B}\right)^{\widetilde{m}_B+n} \exp\left(-\frac{2^{R_s}(1+\gamma_E) -1}{\widetilde{m}_B ^{-1}\widetilde{\overline{\gamma}}_B}\right)}{\Gamma\left(\widetilde{m}_B + n +1\right)}\notag \\
%==================================================================-
%&= \sum_{n = 0}^{+\infty}\frac{\Gamma(\widetilde{m}_B)\left(\frac{\widetilde{m}_B}{\widetilde{\overline{\gamma}}_B}\right)^{\widetilde{m}_B+n} e^{-\frac{2^{R_s}(1+\gamma_E) -1}{\widetilde{m}_B ^{-1}\widetilde{\overline{\gamma}}_B}}}{\Gamma(\widetilde{m}_B + n+1)}\left({2^{R_s}\gamma_E} + {2^{R_s}-1}\right)^{\widetilde{m}_B+n}\notag\\
%-----------------------------------------------------------------
&= \sum_{n = 0}^{+\infty}\frac{\Gamma(\widetilde{m}_B) (2^{R_s}\gamma_E)^{\widetilde{m}_B+n}\left(1 + \frac{2^{R_s}-1}{2^{R_s}\gamma_E}\right)^{\widetilde{m}_B+n}}
{\left(\frac{\widetilde{\overline{\gamma}}_B}{\widetilde{m}_B}\right)^{\widetilde{m}_B+n}
\exp\left(\frac{2^{R_s}(1+\gamma_E) -1}{\widetilde{m}_B ^{-1}\widetilde{\overline{\gamma}}_B}\right)
\Gamma(\widetilde{m}_B + n+1)}\notag\\
%-------------------------------------------------------------------
&= \sum_{n = 0}^{+\infty}\frac{\Gamma(\widetilde{m}_B) \exp\left(-\frac{2^{R_s}(1+\gamma_E) -1}{\widetilde{m}_B ^{-1}\widetilde{\overline{\gamma}}_B}\right)(2^{R_s}\gamma_E)^{\widetilde{m}_B+n}}{\left(\frac{\widetilde{\overline{\gamma}}_B}{\widetilde{m}_B}\right)^{\widetilde{m}_B+n} \Gamma(\widetilde{m}_B + n+1)}\notag \\
&~~~~~~~~\times \sum_{l=0}^{+\infty}\binom{\widetilde{m}_B + n}{l}\left(\frac{2^{R_s}-1}{2^{R_s}\gamma_E}\right)^l,
\end{align}
in which the identity \cite[Eq. (1.110)]{gradshteuin2007table}
\begin{align}
\left(1+\mu\right)^{\alpha} = \sum_{l=0}^{+\infty}\binom{\alpha}{l}\mu^l
\end{align}
is employed.
Substituting \eqref{pdf_SNR_E} and \eqref{gamma_Expand_Exact} into \eqref{outage_probability_1}, we have
\begin{align}\label{outage_probability_2}
&\mathcal{O}\left(R_{s}\right)
= \int_0^{\infty} \left(\frac{\widetilde{m}_E}{\widetilde{\overline{\gamma}}_E}\right)^{N_E\widetilde{m}_E} \frac{\gamma_E^{N_E\widetilde{m}_E-1}}{\Gamma(N_E\widetilde{m}_E)}
\exp\left(\frac{-\widetilde{m}_E \gamma_E}{\widetilde{\overline{\gamma}}_E}\right) \times \notag\\
%-------------------------------------------------------------------
&~~~~~~~~~~~~\sum_{n = 0}^{+\infty}\frac{\exp\left(\!-\!\frac{2^{R_s}(1+\gamma_E) -1}{\widetilde{m}_B ^{-1}\widetilde{\overline{\gamma}}_B}\right)(2^{R_s}\gamma_E)^{\widetilde{m}_B+n}}{\left(\frac{\widetilde{\overline{\gamma}}_B}{\widetilde{m}_B}\right)^{\widetilde{m}_B+n} \Gamma(\widetilde{m}_B + n+1)}\times \notag \\
%-------------------------------------------------------------------
&~~~~~~~~~~~~\sum_{l=0}^{+\infty}\binom{\widetilde{m}_B + n}{l}\left(\frac{2^{R_s}-1}{2^{R_s}\gamma_E}\right)^l   d\gamma_E\notag\\
%%%%%%%%%%%%%%%%%%%%%%%%%%%%%%%%%%%%%%%%%%%%%%%%%%%%%%%%%%%%%%%%%%%%%
&=\frac{ \widetilde{m}_B^{\widetilde{m}_B}\widetilde{m}_E^{N_E\widetilde{m}_E}2^{\widetilde{m}_B R_s}}{\Gamma(N_E\widetilde{m}_E)\widetilde{\overline{\gamma}}_B^{\widetilde{m}_B}
\widetilde{\overline{\gamma}}_E^{N_E\widetilde{m}_E}}
%------
\sum_{n = 0}^{+\infty}\frac{\widetilde{m}_B^n 2^{n R_s}\exp\!\left(\!-\!\frac{\widetilde{m}_B \left(2^{R_s}-1\right)\!}{\widetilde{\overline{\gamma}}_B}\right)}
{\widetilde{\overline{\gamma}}_B^n \Gamma(\widetilde{m}_B +n+1)}\notag\\
%===================================================================
&~~~~\sum_{l = 0}^{+\infty}\frac{\binom{\widetilde{m}_B\!+\!n}{l}\left({2^{R_s}\!\!-\!\!1}\right)^l}
{\!{2^{l R_s}}\!}
\!\!{\int_0^{\infty}}\!\! \frac{\gamma_E^{\widetilde{m}_B \!+\! N_E\widetilde{m}_E \!+\! n\!-\!l\!-\!1}}{\!{\exp\!\left(\frac{\left(2^{R_s}\widetilde{m}_B \widetilde{\overline{\gamma}}_E +  \widetilde{m}_E \widetilde{\overline{\gamma}}_B\right) \gamma_E}{\widetilde{\overline{\gamma}}_B \widetilde{\overline{\gamma}}_E}\right)\!}\!} d \gamma_E.
%%%%%%%%%%%%%%%%%%%%%%%%%%%%%%%%%%%%%%%%%%%%%%%%%%%%%%%%%%%%%%%%%%%%%%
\end{align}
We then obtain the desirable result in \eqref{secercy_outage_final} by solving the integral in \eqref{outage_probability_2} as per the
following identity \cite[Eq. (3.381.4)]{gradshteuin2007table}
\begin{align}\label{integral_2}
\int_0^{\infty} t^{\nu-1} e^{-\mu t} dt = \frac{1}{\mu^{\nu}}\Gamma_G(\nu).
\end{align}
\end{IEEEproof}

We first note the secrecy outage probability derived in \eqref{secercy_outage_final} is a function of Bob and Eve's locations and the beamformer $\mathbf{b}$, all of which are embedded in the parameters $\widetilde{m}_B$, $\widetilde{m}_E$, $\widetilde{\overline{\gamma}}_B$, and $\widetilde{\overline{\gamma}}_E$.
We also note that \eqref{secercy_outage_final} is valid for arbitrary $\widetilde{m}_B$ and $\widetilde{m}_E$ ($\widetilde{m}_B$ and $\widetilde{m}_E$ can be equal), and thus \eqref{secercy_outage_final} is valid for arbitrary $K_B$ and $K_E$. As such, our derived expression for the secrecy outage probability is of more generality than that presented in \cite{yang2013transmit}, which is only valid for integral $\widetilde{m}_B$ and $\widetilde{m}_E$. Although the expression presented in \eqref{secercy_outage_final} involves two infinite series, they both can be approximated by finite series accurately. We approximate the infinite series $\sum_{n=0}^{+\infty}$ and $\sum_{l=0}^{+\infty}$ by truncating them at finite numbers. As we will show in Section~\ref{sec_numerical}, the accuracy of such approximations is acceptable as long as the truncating numbers are larger than approximately 100.
%[???]

An important performance parameter associated with the secrecy outage probability is the secrecy diversity order, which determines the slope of the curve for the secrecy outage probability (in dB) versus $\overline{\gamma}_B$ (in dB) as $\overline{\gamma}_B \rightarrow \infty$ for finite $\overline{\gamma}_E$. Mathematically, the secrecy diversity order is defined as
\begin{align}
\beta = \lim_{\overline{\gamma}_B \rightarrow \infty} \frac{\log_{10} \mathcal{O}\left(R_{s}\right)}{\log_{10} (1/\overline{\gamma}_B)}.
\end{align}
The secrecy diversity order of the LBB scheme is presented in the following corollary.
\begin{corollary}\label{corollary1}
The secrecy diversity order of the LBB scheme is $\widetilde{m}_B$.
\end{corollary}

Following a procedure similar to that used in deriving the secrecy diversity order of the antenna selection schemes presented in \cite{yang2013transmit,yan2014transmit}, we can obtain in a straightforward manner the secrecy diversity order of the LBB scheme as $\widetilde{m}_B$. As such, we omit the proof of the above corollary here. We note that maximum value of $\widetilde{m}_B$ is $(N_A K_B +1)^2/(2 N_A K_B +1)$ due to $|\mathbf{h}_o \mathbf{b}|^2 \leq \|\mathbf{h}_o\|^2 \|\mathbf{b}\|^2 = N_A$.

The probability of non-zero secrecy capacity is defined as the probability that a positive secrecy capacity is achieved. As per \eqref{secrecy_capacity}, it can be formulated as
\begin{align}\label{non-zero}
P_{non} &= \Pr(C_s > 0)\notag \\
&= 1- \int_0^{\infty}f_{\gamma_E}(\gamma_E)\left(\int_{0}^{\gamma_E}f_{\gamma_B}(\gamma_B)d\gamma_B\right) d\gamma_E.
\end{align}
Then, the probability of non-zero secrecy capacity of the LBB scheme is presented in the following corollary.

\begin{corollary}\label{corollary2}
The probability of non-zero secrecy capacity of the LBB scheme is given by
\begin{align}\label{non-zero_los}
P_{non} &= 1 - \frac{\widetilde{m}_B^{\widetilde{m}_B}\widetilde{m}_E^{N_E\widetilde{m}_E}}{\Gamma(N_E \widetilde{m}_E) \widetilde{\overline{\gamma}}_E^{-\widetilde{m}_B}\widetilde{\overline{\gamma}}_B^{-N_E \widetilde{m}_E}}
\sum_{n = 0}^{+\infty}\frac{\widetilde{m}_B^{n} \widetilde{\overline{\gamma}}_E^n}{\Gamma(\widetilde{m}_B + n+1)} \notag\\
%===================================================================
&~~~~~~~~~~~~\times
\frac{ \Gamma\left( \widetilde{m}_B+N_E\widetilde{m}_E +n\right)}{\left(\widetilde{m}_B \widetilde{\overline{\gamma}}_E+\widetilde{m}_E \widetilde{\overline{\gamma}}_B\right)^{\widetilde{m}_B +N_E \widetilde{m}_E +n}}.
\end{align}
\end{corollary}

\begin{IEEEproof}
As per \eqref{non-zero}, the probability of non-zero secrecy capacity can also be formulated as
\begin{align}
P_{non} = 1 - \mathcal{O}(R_s = 0).
\end{align}
Substituting $R_s = 0$ into \eqref{secercy_outage_final}, we obtain the desirable result in \eqref{non-zero_los}.
\end{IEEEproof}

We note that the expression for the probability of non-zero secrecy capacity is simpler than that for the secrecy outage probability and it only involves one infinite series. This infinite series can also be approximated by truncating it at a finite number. This approximation is very accurate even when the truncating number is small (e.g., $10$).

\subsection{Optimal Location-based Beamformer}

A location-based beamformer  can be written as
\begin{align}\label{los_B}
\mathbf{b} = \frac{1}{\sqrt{N_A}}\left[1,\cdots,\exp(-j (N_A -1)\tau_A \cos \psi)\right]^T,
\end{align}
where $\psi$ ($0 \leq\psi \leq \pi$) is the beamforming direction. In this work we define the optimal location-based beamformer, $\mathbf{b}^{\ast}$, as the one that minimizes the secrecy outage probability for a given $R_s$.
Therefore, defining
\begin{align}\label{optimal_Angle}
\psi^{\ast} = \argmin_{0 \leq \psi \leq \pi} \mathcal{O}(R_s),
\end{align}
%\end{definition}
and
setting $\psi=\psi^{\ast}$ in \eqref{los_B} completely define the optimal beamformer $\mathbf{b}^{\ast}$. We note that the value range of $\psi$ is selected based on the symmetric property of the ULA (e.g., $\psi=\pi/3$ and $\psi=-\pi/3$ lead to the same beamformer $\mathbf{b}$).
We note that \eqref{optimal_Angle} is a one-dimensional optimization problem, which can be solved through numerical search. Substituting $\mathbf{b}^{\ast}$ into \eqref{secercy_outage_final}, we achieve the minimum secrecy outage probability of the  LBB scheme, which is denoted as $\mathcal{O}^{\ast}(R_s)$. We would like to highlight that $\psi^{\ast}$ can be analytically determined in some special cases as detailed in the following corollaries.

\begin{corollary}\label{coro3}
For $K_B > 0$, the solution to \eqref{optimal_Angle} is $\psi^{\ast} = \theta_B$ in the following cases: (i) when $\overline{\gamma}_B \rightarrow \infty$ for finite $\overline{\gamma}_E$, (ii) when $K_E = 0$, or (iii) when $\theta_E$ is unavailable at Alice.
\end{corollary}
%[??? I am afraid this proof is unreadable and likely wrong as a formal proof - unless you can remove the words "mainly" and "intend" from this and then try and formally prove each (separately) of cases (i)-(iii) above.
\begin{IEEEproof}
In Case (i), as $\overline{\gamma}_B \rightarrow \infty$ the secrecy diversity order determines the secrecy outage probability. As such, as $\overline{\gamma}_B \rightarrow \infty$ the optimal location-based beamformer is to maximize the secrecy diversity order given in Corollary~\ref{corollary1} (i.e., $\widetilde{m}_B$) in order to minimize the secrecy outage probability. To this end, $\psi^{\ast}$ is to maximize $\widetilde{K}_B$. Following \eqref{widetilde_KB}, $\psi^{\ast}$ finally is to maximize $|\mathbf{h}_o\mathbf{b}|^2$. In Case (ii), there is no LOS component in the eavesdropper's channel due to $K_E = 0$ and $\psi$ does not impact $\gamma_E$. As such, $\psi^{\ast}$ is to maximize $\gamma_B$ in order to minimize the secrecy outage probability. Following \eqref{SNR_B}, $\psi^{\ast}$ finally is to maximize $|\mathbf{h}_o\mathbf{b}|^2$ in this case. In Case (iii), Alice is not sure how $\psi$ impacts $\gamma_E$ since $\theta_E$ is unknown. Then, $\psi$ is to maximize $\gamma_B$ and thus to maximize $|\mathbf{h}_o\mathbf{b}|^2$ based on \eqref{SNR_B}.

\begin{figure}[!t]
    \begin{center}
   {\includegraphics[width=3.5in, height=2.9in]{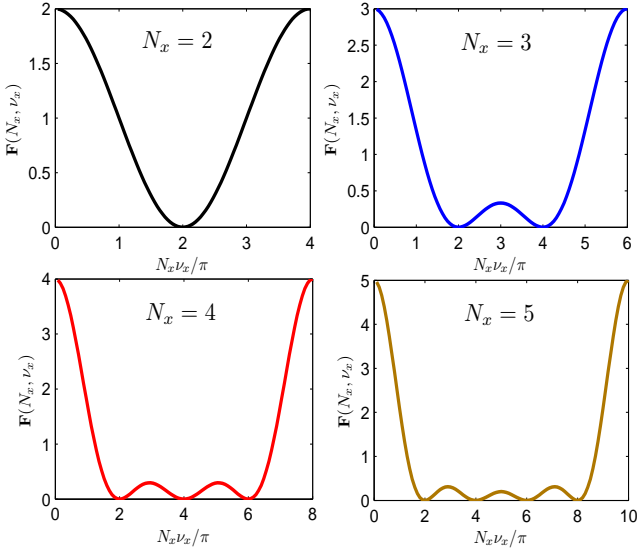}}
    \caption{$\mathbf{F}(N_x, \nu_x)$ versus $N_x \nu_x/\pi$ for different values of $N_x$.}\label{fig:F_function}
    \end{center}
\end{figure}

As we can see from the above discussion, in all three cases of the corollary the value of $\psi^{\ast}$ is  the one that maximizes $|\mathbf{h}_o \mathbf{b}|^2$. So, to complete the proof we now prove that this value is indeed $\theta_B$.
Denoting $\nu_A = \tau_A (\cos \theta_B - \cos \psi)$, as per \eqref{h_o_definition} and \eqref{los_B}, for $\nu_A \neq 0$ we have
\begin{align}\label{hb_1}
\mathbf{h}_o \mathbf{b} &= \frac{1}{\sqrt{N_A}} \frac{\exp \left(j N_t \nu_A \right) - 1}{\exp \left(j \nu_A \right) - 1}\notag \\
&= \frac{1}{\sqrt{N_A}}\frac{-e^{j N_A \nu_A /2}\left(-e^{-j N_A \nu_A /2} -e^{jN_A \nu_A/2}\right)}{-e^{j \nu_A/2}\left(-e^{-j \nu_A/2} -e^{j \nu_A /2}\right)}\notag\\
&= \frac{1}{\sqrt{N_A}} \frac{\sin \left(\frac{1}{2}N_A \nu_A\right)}{\sin \left(\frac{1}{2} \nu_A\right)}e^{j \nu_A (N_A-1)/2}.
\end{align}
For $\nu_A = 0$, we have $\mathbf{h}_o \mathbf{b} = \sqrt{N_A}$.
Then, following \eqref{hb_1} we have
\begin{align}
|\mathbf{h}_o\mathbf{b}|^2 = \mathbf{F}(N_A, \nu_A),
\end{align}
where $\mathbf{F}(\cdot, \cdot)$ is defined as
\begin{align}\label{hb}
\mathbf{F}(N_x, \nu_x) =\left\{
\begin{array}{ll}
N_x, &  \nu_x = 0,\\
\frac{1}{{N_x}}\left(\frac{\sin \left(\frac{1}{2}N_x \nu_x \right)}{\sin \left(\frac{1}{2} \nu_x\right)}\right)^2, & 0 \leq \nu_x < 2 \pi.
\end{array}
\right.
\end{align}
It is straightforward to prove that the maximum value of $\mathbf{F}(N_x, \nu_x)$ is $N_x$, which is achieved for $\nu_x = 0$. This is also confirmed by Fig.~\ref{fig:F_function}, where we plot $\mathbf{F}(N_x, \nu_x)$ versus $N_x \nu_x/\pi$ for different value of $N_x$. As such, $|\mathbf{h}_o \mathbf{b}|^2$ is maximized when $\nu_A = 0$ and thus we have $\psi^{\ast} = \theta_B$ (we ignore the negative solutions due to $0 \leq \psi \leq \pi$) in order to maximize $|\mathbf{h}_o \mathbf{b}|^2$.
\end{IEEEproof}

We note that for $\psi^{\ast} = \theta_B$ we have $\mathbf{b}^{\ast} = \mathbf{h}_o^{\dag}/\sqrt{N_A}$ and $|\mathbf{h}_o\mathbf{b}|^2 = N_A$. As such, we have $\widetilde{K}_B = N_A K_B$ and $\widetilde{\overline{\gamma}}_B = {(N_A K_B + 1)\overline{\gamma}_B}/ (1+K_B)$. We denote the secrecy outage probability of the LBB scheme with unknown Eve's location (i.e., $\psi^{\ast} = \theta_B$) as $\mathcal{O}_b(R_s)$.

\begin{corollary}\label{coro4}
For $K_E > 0$, the (multiple) solution to \eqref{optimal_Angle} is $\psi^{\ast} = \arccos \left(\cos \theta_E + \frac{2 n_A \pi}{N_A \tau_A}\right)$, $n_A = 1, \dots, N_A -1$, in the following cases: (i) when $\overline{\gamma}_E \rightarrow \infty$ for finite $\overline{\gamma}_B$, (ii) when $K_B = 0$, or (iii) when $\theta_B$ is unavailable at Alice.
\end{corollary}

\begin{IEEEproof}
Following similar arguments to those used in the proof of Corollary~\ref{coro3}, we know that $\psi^{\ast}$ is to minimize $|\mathbf{g}_o\mathbf{b}|^2$ for all three cases in Corollary~\ref{coro4}. The value of $|\mathbf{g}_o\mathbf{b}|^2$ is given by
\begin{align}
|\mathbf{g}_o\mathbf{b}|^2 = \mathbf{F}(N_A, \nu_E),
\end{align}
where $\nu_E = \tau_A (\cos \theta_E - \cos \psi)$. We note that the minimum value of $\mathbf{F}(N_x, \nu_x)$ is achieved when $\nu_x = 2 n_x \pi$ for $n_x = 1, \dots, N_x-1$, which is also confirmed by Fig.~\ref{fig:F_function}. As such, $|\mathbf{g}_o\mathbf{b}|^2$ is minimized when $\nu_E = 2 n_A \pi$ for $n_A = 1, \dots, N_A-1$, and thus we obtain Corollary~\ref{coro4}.
\end{IEEEproof}

\section{Numerical Results}\label{sec_numerical}

In this section we  present numerical simulations to verify our secrecy performance analysis of the LBB scheme, and examine the impact of different system parameters (e.g., $K_B$, $K_E$, $\overline{\gamma}_B$, and $\overline{\gamma}_E$) on the LBB scheme. To better illustrate the gains obtained by our scheme, we will also present simulations of the secrecy performance of the NB (non-beamforming) scheme. This latter scheme represents the case when an isotropic beamforming pattern is produced by Alice (see Appendix~A for an analytical analysis of this scheme). To conduct simulations, we deploy Bob and Eve at specific locations and then map such locations into $\overline{\gamma}_B$ and $\overline{\gamma}_E$, respectively. Such a mapping is based on Alice's transmit power (i.e., $P$) and path loss exponents of the main channel and the eavesdropper's channel (i.e., $\eta_B$ and $\eta_E$). For presentation convenience, we only specify the values of $\overline{\gamma}_B$ and $\overline{\gamma}_E$ adopted in our following simulations. We note that in the following figures we use ``Theo'' and ``Simu'' as the abbreviations of ``Theoretic'' and ``Simulated'', respectively.

\begin{figure}[!t]
    \begin{center}
   {\includegraphics[width=3.5in, height=2.9in]{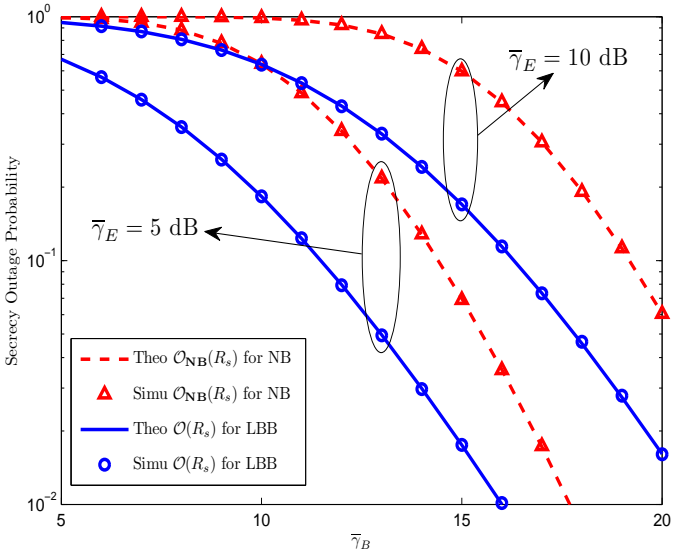}}
    \caption{Secrecy outage probabilities under Nakagami channels versus different values of $\overline{\gamma}_B$, where $m_B = 1.35, m_E = 1.33, \overline{\lambda}_0 = 0.85, N_A = 3, N_E = 2, \text{and}~R_s = 1$.}\label{fig:Nakagami_check}
    \end{center}
\end{figure}

In Fig.~\ref{fig:Nakagami_check} we first verify our derived secrecy outage probabilities for Nakagami fading channels. To this end, we generate channel realizations as per the Nakagami fading channel, where we have set $\widetilde{m}_B =2 m_B$, $\widetilde{m}_E  = m_E$, $\widetilde{\overline{\gamma}}_B = 3 \overline{\gamma}_B$, and $\widetilde{\overline{\gamma}}_E  = \overline{\gamma}_E$, where $m_B = (K_B +1)^2/(2 K_B +1)$ and $m_E = (K_E +1)^2/(2 K_E +1)$. The theoretic
secrecy outage probability of the LBB scheme, $\mathcal{O}(R_s)$, and the secrecy outage probability of
the NB scheme, denoted as $\mathcal{O}_{\mathbf{NB}}(R_s)$, are obtained through \eqref{secercy_outage_final} and \eqref{outage_probability_NB}, respectively, where relevant infinite series are truncated at 100. In this figure, we observe that the theoretic $\mathcal{O}(R_s)$ and $\mathcal{O}_{\mathbf{NB}}(R_s)$ precisely match the simulated $\mathcal{O}(R_s)$ and $\mathcal{O}_{\mathbf{NB}}(R_s)$, respectively. This confirms the correctness of our derived secrecy outage probabilities.

\begin{figure}[!t]
    \begin{center}
   {\includegraphics[width=3.5in, height=2.9in]{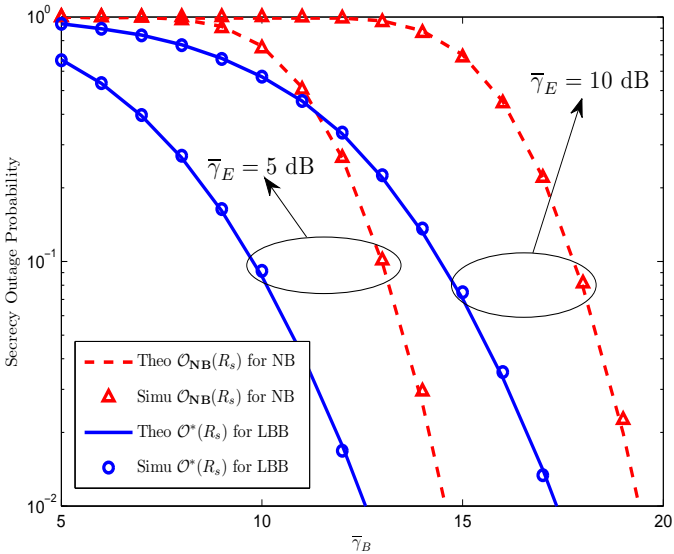}}
    \caption{Secrecy outage probabilities under Rician channels versus different values of $\overline{\gamma}_B$, where $N_A = 3, N_E = 2, K_B = 10 ~\text{dB}, K_E = 5 ~\text{dB}, \theta_B = \pi/3, \theta_E = \pi/4, \text{and}~R_s = 1$.}\label{fig:Pout_verify}
    \end{center}
\end{figure}

Recall that for mathematical convenience, our analysis  approximates a Rician channel with a Nakagami channel. To see the effect of this, in
Fig.~\ref{fig:Pout_verify} we again plot the secrecy outage probabilities of the LBB scheme and the NB scheme, but this time for  specific Rician fading channels.
In this figure, we  observe that the simulated minimum secrecy outage probability of the LBB scheme, $\mathcal{O}^{\ast}(R_s)$, and the secrecy outage probability of the NB scheme, $\mathcal{O}_{\mathbf{NB}}(R_s)$,  match extremely well the theoretic $\mathcal{O}^{\ast}(R_s)$ and $\mathcal{O}_{\mathbf{NB}}(R_s)$, respectively, thus confirming the validity of our channel approximation.
We note that we have set $\theta_E$ very close to $\theta_B$ in Fig.~\ref{fig:Pout_verify} (i.e., $\theta_B = \pi/3$ and $\theta_E = \pi/4$). The gap between $\mathcal{O}^{\ast}(R_s)$ and $\mathcal{O}_{\mathbf{NB}}(R_s)$ can even be larger when $\theta_E$ is not so close to $\theta_B$.

\begin{figure}[!t]
    \begin{center}
   {\includegraphics[width=3.5in, height=2.9in]{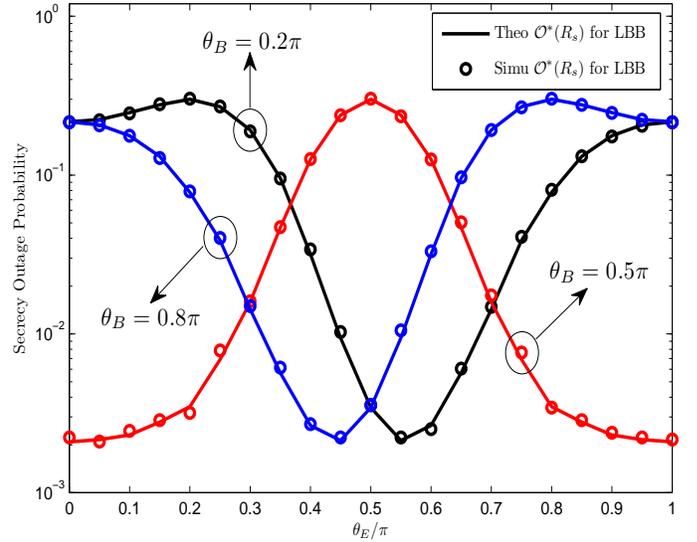}}
    \caption{Minimum secrecy outage probability of the LBB scheme versus different values of $\theta_E$, where $N_A = 2, N_E = 2, K_B = 10~\text{dB}, K_E = 10~\text{dB}, \overline{\gamma}_B = 10~\text{dB}, \overline{\gamma}_E = 10~\text{dB}, \text{and}~R_s = 1$.}\label{fig:optimal_psi}
    \end{center}
\end{figure}

In Fig.~\ref{fig:optimal_psi}, we plot the minimum secrecy outage probability of the LBB scheme, $\mathcal{O}^{\ast}(R_s)$, versus different values of $\theta_E$.
Again we observe that the theoretic $\mathcal{O}^{\ast}(R_s)$  matches extremely well the simulated $\mathcal{O}^{\ast}(R_s)$, which again confirms the validity of our analysis. Fig.~\ref{fig:optimal_psi} is also useful in that it more visually represents how the minimum secrecy outage probability of the LBB scheme depends on the locations of Bob and Eve. For example, $\mathcal{O}^{\ast}(R_s)$ is maximized when $\theta_B = \theta_E$.
In the simulations to obtain Fig.~\ref{fig:optimal_psi}, we also observe that the optimal beamforming direction $\psi^{\ast}$ shifts away from $\theta_B$ as $\theta_E$ approaches to $\theta_B$.

\begin{figure}[!t]
    \begin{center}
   {\includegraphics[width=3.5in, height=2.9in]{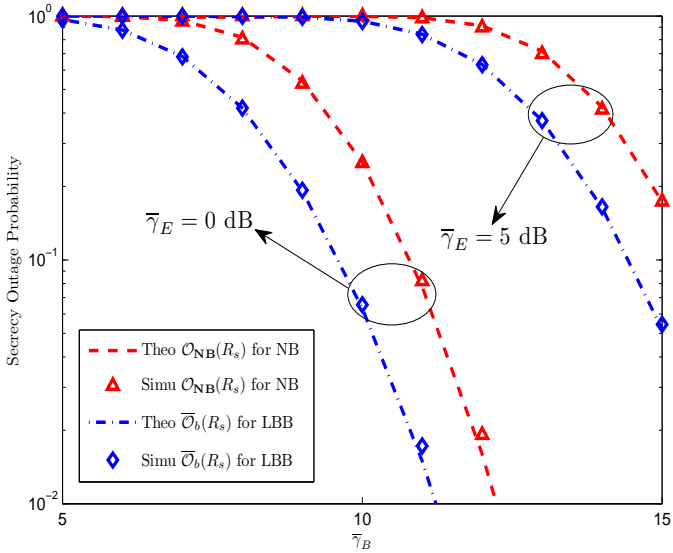}}
    \caption{Secrecy outage probabilities without Eve's location versus different values of $\overline{\gamma}_B$, where $N_A = 3, N_E = 4, K_B = 10 ~\text{dB}, K_E = 5 ~\text{dB}, \theta_B = \pi/3, \text{and}~R_s = 1$.}\label{fig:Pout_noEve}
    \end{center}
\end{figure}

In Fig.~\ref{fig:Pout_noEve}, we examine the secrecy outage probability of the LBB scheme without knowing Eve's location, $\mathcal{O}_b(R_s)$. As per Corollary~3, we know that $\mathbf{b}^{\ast} = \mathbf{h}^{\dag}/\|\mathbf{h}\|$ when Eve's location is unavailable at Alice.
%rrdel We note that the secrecy outage probability of the LBB scheme for $\mathbf{b}^{\ast} = \mathbf{h}^{\dag}/\|\mathbf{h}\|$, $\mathcal{O}_b(R_s)$, is still a function of $\theta_E$.
In Fig.~\ref{fig:Pout_noEve} we also compare the the solution with no information on Eve's location to the NB scheme.
To conduct a fair comparison with the NB scheme, we assume Eve's location is uniformly distributed on a circle centered at Alice, i.e., $\theta_E$ uniformly distributes between $0$ and $2 \pi$, $\theta_E \sim \mathcal{U}[0, 2\pi]$. We then average $\mathcal{O}_b(R_s)$ over $\theta_E$ to obtain the average secrecy outage probability, denoted as $\overline{\mathcal{O}}_b(R_s)$.
As expected, we observe that $\overline{\mathcal{O}}_b(R_s)$ is lower than $\mathcal{O}_{\mathbf{NB}}(R_s)$, which demonstrates that the LBB scheme still outperforms the NB scheme \emph{on average}, even when Eve's location is unavailable at Alice. This is due to the fact that the LBB scheme improves the quality of the main channel based on Bob's location, which on average reduces the secrecy outage probability. However, the most important result obtained from the simulations of Fig.~\ref{fig:Pout_noEve} is that the secrecy outage probability of the LBB scheme without Eve's location increases (e.g., by approximately a factor of 5 for $\overline{\gamma}_B = 10$dB) relative to that of the LBB scheme with Eve's location. This quantifies the value of the location information of Eve to the beamformer solution.

It is worth mentioning how relaxations of some key assumptions we have made impact the results presented here. Of course, in reality it will never be the case that all reported locations, all $K$ map information, and all path loss exponents are known with zero error. Errors in these quantities are intermingled in the sense that an error in one leads to an error in another. We have attempted to encompass such correlated  errors in a range of additional simulations. Our general result is that a percentage error of $15\%$ in any of these inputs leads to an approximately $10\%$ percentage error in our reported outage probabilities. For anticipated error inputs, we can therefore say that our analysis remains reasonably accurate.

Finally, although outside the spirit of our low-complexity LBB scheme, it is perhaps worth discussing the gains to be made when the full CSI information of the main channel is made available to Alice (where the transmission scheme is named as the full-CSI scheme).  If Eve's location is also available at Alice in the full-CSI scheme, the full-CSI scheme will of course outperform the LBB scheme for any values of system parameters. For example, under the simulation settings of Fig.~\ref{fig:Pout_verify}, the secrecy outage probability of this full-CSI scheme with Eve's location is $15\%$ lower than that of the LBB scheme for $\overline{\gamma}_B = 10$dB and $\overline{\gamma}_E = 5$dB (determined from simulations). If Eve'e location is unavailable at Alice (in both schemes) then the full-CSI scheme outperforms the LBB scheme by $40\%$ for $\overline{\gamma}_B = 10$dB and $\overline{\gamma}_E = 0$dB under the same simulation setting of Fig.~\ref{fig:Pout_noEve} (determined from simulations and analysis).  For completeness, the secrecy performance analysis of the full-CSI scheme is given in Appendix~B.

\section{Conclusions}\label{sec_conclusion}

We proposed and analyzed a novel beamforming scheme in the wiretap channel where both the main channel and the eavesdropper's channel are subject to Rician fading. Our new LBB scheme solely requires as inputs the location information of Bob and Eve, and does not require the CSI of the main channel or the eavesdropper's channel. We  derived the secrecy outage probability of the LBB scheme in a closed-form expression valid for arbitrary values of $K_B$ and $K_E$. We then determined the optimal location-based beamformer that minimizes the secrecy outage probability. Comparisons with a range of other schemes were then carried out so as to better understand the performance gains offered by our location-based solution. The work we presented will be important for a range of application scenarios in which Rician channels are expected to be dominant and where location information of potential users and adversaries are known.

\begin{appendices}

\section{Secrecy Performance of the NB Scheme}\label{sec_NB}

In the NB scheme, Alice distributes her total transmit power uniformly among the $N_A$ orthogonal independent transmit directions (i.e., the covariance matrix of $\mathbf{b} x$ is $P \mathbf{I}_{N_A}/N_A$) \cite{telatar1999capacity,anna2009outage}. Then, the SNR at Bob is given by \cite{telatar1999capacity,anna2009outage}
\begin{align}
\gamma_B^{\mathbf{NB}} &= \frac{\overline{\gamma}_B||\mathbf{h}||^2}{N_A}.
\end{align}
Interpreting  Rician fading as a special case of  Nakagami fading, the pdf of $\gamma_B^{\mathbf{NB}}$ can be approximated by
\begin{align}
f_{\gamma_B^{\mathbf{NB}}}(\gamma) = \frac{m_B^{N_A m_B}\gamma^{N_A m_B -1} e^{-\frac{N_A m_B \gamma}{\overline{\gamma}_B}}}{\Gamma(N_A m_B)(\overline{\gamma}_B/N_A)^{N_A m_B}}.
\end{align}
We assume that Eve applies MRC to combine the received signals at different antenna elements. As such, the SNR at Eve is given by
\begin{align}
\gamma_E^{\mathbf{NB}} = \frac{\overline{\gamma}_E||\mathbf{s}_0^{\dag}\mathbf{G}||^2}{N_A} = \frac{\overline{\gamma}_E \lambda_0^2}{N_A},
\end{align}
where $\mathbf{s}_0$ is the $N_E \times 1$ eigenvector for the largest eigenvalue $\lambda_0$ of $\mathbf{G}$. The theoretical expression for the distribution of $\lambda_0^2$ has been derived in \cite{kang2003largest}. However, this expression is too complicated to be used for further analysis. To make progress, we adopt the simple approximation for the pdf of $\lambda_0^2$ proposed in \cite{tani2007approx}. As such, the pdf of $\gamma_E^{\mathbf{NB}}$ can be approximated by
\begin{align}\label{pdf_snr_E_uspa}
f_{\gamma_E^{\mathbf{NB}}}(\gamma) = \frac{(N_A m_E)^{N_A N_E m_E}\gamma^{N_A N_E m_E -1}}{\Gamma(N_A N_E m_E)(\overline{\gamma}_E \overline{\lambda}_0)^{N_A N_E m_E}}\exp\left(\!-\!\frac{N_A m_E \gamma}{\overline{\gamma}_E\overline{\lambda}_0}\right),
\end{align}
where $\overline{\lambda}_0$ is the mean of the per-branch largest eigenvalue (i.e., $\overline{\lambda}_0 = \mathbb{E}[\lambda_0]/N_A N_E$). The value of $\overline{\lambda}_0$ can be approximated by\cite{tani2007approx}
\begin{equation}\label{lambda_Approximation}
\overline{\lambda}_0=\left\{
\begin{array}{ll}
\frac{K_E}{K_E + 1} + \frac{1}{K_E + 1} \frac{N_A + N_E}{N_A N_E +1}~\;, &  K_E \geq 0.5,\\
\left(\frac{N_A + N_E}{N_A N_E + 1}\right)^{\frac{4-K_E}{6}}~\;, & K_E < 0.5.\;
\end{array}
\right.
\end{equation}
We note that we have $\overline{\lambda}_0 = 1$ for arbitrary $K_E$ when $N_E = 1$.

%As such, \eqref{pdf_snr_E_uspa} for $N_E = 1$ will be the pdf of the SNR for the $1 \times N_A$ Nakagami fading channel.

Following a similar procedure to that used in deriving $\mathcal{O}(R_s)$ in Theorem~\ref{theorem1}, the secrecy outage probability of the NB scheme is derived as
\begin{align}\label{outage_probability_NB}
&\mathcal{O}_{\mathbf{NB}}\left(R_{s}\right)\!=\!\!\int_0^{\infty}\!\!{f_{\gamma_E^{\mathbf{NB}}}(\gamma_E)}\!
\!\left[\!\int_{0}^{2^{R_s}(1+\gamma_E) \!-\!1}\!f_{\gamma_B^{\mathbf{NB}}}(\gamma_B)d\gamma_B\right]\! d\gamma_E \notag \\
%%%%%%%%%%%%%%%%%%%%%%%%%%%%%%%%%%%%%%%%%%%%%%%%%%%%%%%%%%%%%%%%%%%%
&=\frac{ m_B^{N_A m_B}m_E^{N_A N_E m_E}2^{N_A m_B R_s}}{\Gamma(N_A N_E m_E)\overline{\gamma}_B^{-N_A N_E m_E}
(\overline{\gamma}_E \overline{\lambda}_0)^{-N_A m_B}}\times \notag\\
%------
&~~~\sum_{n = 0}^{+\infty}\frac{ m_B^n 2^{n R_s}\exp\left(-\frac{N_A m_B \left(2^{R_s}-1\right)}{\overline{\gamma}_B}\right)}
{\overline{\gamma}_B^n \Gamma(N_A m_B +n+1)}\times \notag\\
%===================================================================
&~~~\sum_{l = 0}^{+\infty}\frac{\binom{N_A m_B\!+\!n}{l}\left({2^{R_s}\!\!-\!\!1}\right)^l}
{\!{N_A^{-l}2^{l R_s}}\!}\times\notag \\
%===================================================================
&~~~\frac{\left(\overline{\gamma}_B \overline{\gamma}_E \overline{\lambda}_0\right)^{n-l}\Gamma_G(N_A m_B + N_A N_E m_E +n-l)}{\left(2^{R_s}m_B \overline{\gamma}_E \overline{\lambda}_0 + m_E \overline{\gamma}_B \right)^{N_A m_B + N_A N_E m_E +n-l}}.
\end{align}

%\rrdel It is worth to highlight that this is the first time to examine the secrecy performance of the NB scheme for Rican fading channels.
As per \eqref{outage_probability_NB}, we can see that the secrecy outage probability of the NB scheme is independent of $\theta_B$ and $\theta_E$. However, \eqref{outage_probability_NB} is a function of $\overline{\gamma}_B$ and $\overline{\gamma}_E$, which are dependent on $d_B$ and $d_E$, respectively.
We note that the secrecy diversity order of the NB scheme is $N_A m_B$, which is the full secrecy diversity order.
Also, following a similar procedure to that used in deriving $P_{non}$ in Corollary~2, the probability of non-zero secrecy capacity of the NB scheme is derived as
\begin{align}\label{non-zero_los_uspa}
\begin{split}
P_{non}^{\mathbf{NB}} &= 1 - \frac{m_B^{N_A m_B}m_E^{N_A N_E m_E}\overline{\gamma}_B^{N_A N_E m_E}}{\Gamma(N_A N_E m_E) (\overline{\gamma}_E \overline{\lambda}_0)^{-N_A m_B}} \times \\
%===================================================================
&~~~~~~~~\sum_{n = 0}^{+\infty}\frac{m_B^{n}\left(\overline{\gamma}_E \overline{\lambda}_0\right)^n}{\Gamma(N_A m_B + n+1)} \times \\
%===================================================================
&~~~~~~~~
\frac{ \Gamma\left( N_A m_B+N_A N_Em_E +n\right)}{\left(m_B \overline{\gamma}_E \overline{\lambda}_0+ m_E \overline{\gamma}_B\right)^{N_A m_B + N_A N_E m_E +n}}.
\end{split}
\end{align}

\section{Secrecy Performance of the Full-CSI Scheme}\label{sec_MRT}

In the full-CSI scheme, Alice knows the CSI of the main channel (Bob feeds back the CSI to Alice), but Alice does not know the CSI of the eavesdropper's channel or Eve's location.  Then, Alice adopts $\mathbf{h}^{\dag}/\|\mathbf{h}\|$ as the beamformer $\mathbf{b}$ to maximize the SNR of the main channel \cite{big2001limiting,anna2009outage} in order to minimize the secrecy outage probability. The SNR at Bob of the full-CSI scheme is given by \cite{big2001limiting,anna2009outage}
\begin{align}
\gamma_B^{\mathbf{CSI}} = {\overline{\gamma}_B||\mathbf{h}||^2}.
\end{align}
Again using the Nakagami fading to approximate the Rician fading, the pdf of $\gamma_B^{\mathbf{CSI}}$ can be approximated by
\begin{align}
f_{\gamma_B^{\mathbf{CSI}}}(\gamma) = \frac{m_B^{N_A m_B}\gamma^{N_A m_B -1} \exp\left(-\frac{m_B \gamma}{\overline{\gamma}_B}\right)}{\Gamma(N_A m_B)\overline{\gamma}_B^{N_A m_B}}.
\end{align}
We assume that Eve knows $\mathbf{h}$ by eavesdropping on the feedback from Bob to Alice.  We also assume that Eve knows that Alice adopts $\mathbf{h}^{\dag}/\|\mathbf{h}\|$ as the beamformer. Assuming that Eve applies MRC to combine the received signals at different antenna elements, the SNR at Eve is given by
\begin{align}
\gamma_E^{\mathbf{CSI}} = \frac{\overline{\gamma}_E\|\mathbf{G}\mathbf{h}^{\dag}\|^2}
{\|\mathbf{h}\|^2} = \overline{\gamma}_E \sum_{i=1}^{N_E} \gamma_{E,i}^{\mathbf{CSI}},
\end{align}
where $\gamma_{E,i}^{\mathbf{CSI}} = |\mathbf{g}_i \mathbf{h}^{\dag}|/\|\mathbf{h}\|$. In order to derive the pdf of $\gamma_E^{\mathbf{CSI}}$, we next first derive the pdf of $\gamma_{E,i}^{\mathbf{CSI}}$.
As per \eqref{h_definition} and \eqref{g_definition}, we have
\begin{align}
\frac{\mathbf{g}_i\mathbf{h}^{\dag}}{\|\mathbf{h}\|} &= \frac{1}{\|\mathbf{h}\|}\left(e_o \epsilon_i \mathbf{g}_o + e_r \mathbf{g}_r\right)\left(b_o \mathbf{h}_o + b_r \mathbf{h}_r\right)^{\dag} \notag \\
&= \frac{b_o e_o \epsilon_i \mathbf{g}_o \mathbf{h}_o^{\dag}}{\|\mathbf{h}\|} + \frac{b_r e_o \epsilon_i \mathbf{g}_o \mathbf{h}_r^{\dag}}{\|\mathbf{h}\|} + \frac{e_r \mathbf{g}_r \mathbf{h}^{\dag}}{\|\mathbf{h}\|},
\end{align}
where
\begin{align}
b_o = \sqrt{\frac{K_B}{K_B +1}}, ~~b_r = \sqrt{\frac{1}{K_B +1}}, \notag \\
e_o = \sqrt{\frac{K_E}{K_E +1}}, ~~e_r = \sqrt{\frac{1}{K_E +1}}. \notag
\end{align}
To make progress, we make the following approximation
\begin{align}\label{h_m_definition}
\frac{\mathbf{g}_i\mathbf{h}^{\dag}}{\|\mathbf{h}\|} \approx \underbrace{\frac{b_o e_o \epsilon_i \mathbf{g}_o \mathbf{h}_o^{\dag}}{\sqrt{N_A}}}_{h^{\mathbf{CSI}}_o} + \underbrace{\frac{b_r e_o \epsilon_i \mathbf{g}_o \mathbf{h}_r^{\dag}}{\sqrt{N_A}} + \frac{e_r \mathbf{g}_r \mathbf{h}^{\dag}}{\sqrt{N_A}}}_{h^{\mathbf{CSI}}_r}.
\end{align}
We note that in \eqref{h_m_definition} $h^{\mathbf{CSI}}_o$ is deterministic and $h^{\mathbf{CSI}}_r$ is a circularly-symmetric complex Gaussian random variable. As such, ${\mathbf{g}_i\mathbf{h}^{\dag}}/\sqrt{N_A}$ follows a Rician distribution. Following \eqref{h_m_definition}, we have
\begin{align}
|h^{\mathbf{CSI}}_o|^2 &= \frac{b_o^2 e_o^2 |\mathbf{g}_o \mathbf{h}_o^{\dag}|^2}{N_A} = \frac{K_B K_E |\mathbf{g}_o \mathbf{h}_o^{\dag}|^2}{N_A(1+K_B)(1+K_E)},\notag
\end{align}
and
\begin{align}
\mathbb{E}[|h^{\mathbf{CSI}}_r|^2] &= \frac{b_r^2 e_o^2}{{N}}\mathbb{E}[|\mathbf{g}_o \mathbf{h}_r^{\dag}|^2] + \frac{e_r^2}{{N}}\mathbb{E}[|\mathbf{g}_r \mathbf{h}^{\dag}|^2]\notag \\
&= b_r^2 e_o^2 + e_r^2 \notag \\
&= \frac{K_B + K_E + 1}{(K_B+1)(K_E+1)}.\notag
\end{align}
Then, the Rician $K$-factor of ${\mathbf{g}_i\mathbf{h}^{\dag}}/\sqrt{N_A}$ is given by
\begin{align}
\ddot{K}_E \triangleq \frac{|h^{\mathbf{CSI}}_o|^2}{\mathbb{E}[|h^{\mathbf{CSI}}_r|^2]}= \frac{K_B K_E |\mathbf{g}_o \mathbf{h}_o^{\dag}|^2}{N_A(K_B + K_E + 1)}.
\end{align}
Following \eqref{h_m_definition}, we also have
\begin{align}
\ddot{\overline{\gamma}}_E \triangleq \mathbb{E}[\gamma_{Ei}^{\mathbf{CSI}}] &= \overline{\gamma}_E \left(|h^{\mathbf{CSI}}_o|^2 + \mathbb{E}[|h^{\mathbf{CSI}}_r|^2]\right)\notag \\
&= \frac{K_B K_E |\mathbf{g}_o \mathbf{h}_o^{\dag}|^2 \!+\! N_A (K_B \!+\! K_E \!+\! 1)}{\overline{\gamma}_E^{-1} N_A(K_B+1)(K_E+1)}.
\end{align}
Then, the pdf of $\gamma_{Ei}^{\mathbf{CSI}}$ can be approximated by
\begin{align}\label{pdf_SNR_Ei_MRT}
f_{\gamma_{E,i}^{\mathbf{CSI}}}(\gamma) = \left(\frac{\ddot{m}_E}{\ddot{\overline{\gamma}}_E}\right)^{\ddot{m}_E} \frac{\gamma^{\ddot{m}_E-1}}{\Gamma(\ddot{m}_E)}\exp\left(\frac{-\ddot{m}_E \gamma}{\ddot{\overline{\gamma}}_E}\right),
\end{align}
where $\ddot{m}_E = (\ddot{K}_E +1)^2/(2 \ddot{K}_E +1)$. Since $\gamma_{E,i}^{\mathbf{CSI}}$ are independent from each other, the pdf of $\gamma_E^{\mathbf{CSI}}$ can be approximated by
\begin{align}\label{pdf_SNR_E_MRT}
f_{\gamma_E^{\mathbf{CSI}}}(\gamma) = \left(\frac{\ddot{m}_E}{\ddot{\overline{\gamma}}_E}\right)^{N_E \ddot{m}_E} \frac{\gamma^{N_E \ddot{m}_E-1}}{\Gamma(N_E \ddot{m}_E)}\exp\left(\frac{-\ddot{m}_E \gamma}{\ddot{\overline{\gamma}}_E}\right).
\end{align}

Following a similar procedure to that used in deriving $\mathcal{O}(R_s)$ in Theorem~\ref{theorem1}, the secrecy outage probability of the full-CSI scheme is then derived as
\begin{align}\label{secercy_outage_MRT}
&\mathcal{O}_{\mathbf{CSI}} (R_s) \!=\!\!\int_0^{\infty}\!\!{f_{\gamma_E^{\mathbf{CSI}}}(\gamma_E)}\!
\!\left[\!\int_{0}^{2^{R_s}(1+\gamma_E) \!-\!1}\!f_{\gamma_B^{\mathbf{CSI}}}(\gamma_B)d\gamma_B\right]\! \!d\gamma_E\! \notag \\
&=
\frac{ m_B^{N_A m_B}\ddot{m}_E^{N_E \ddot{m}_E}2^{N_A m_B R_s}}{\Gamma(N_E \ddot{m}_E)\overline{\gamma}_B^{-N_E\ddot{m}_E}
\ddot{\overline{\gamma}}_E^{- N_A m_B}}\times \notag \\
%------
&\quad \quad\sum_{n = 0}^{+\infty}\frac{2^{n R_s}\exp\left(\!-\!\frac{m_B \left(2^{R_s}\!-\!1\right)}{\overline{\gamma}_B}\right)}
{m_B^{\!-\!n}\ddot{\overline{\gamma}}_E^{-n} \Gamma(N_A m_B +n+1)} \times \\
%===================================================================
&\sum_{l = 0}^{+\infty}\frac{\binom{N_A m_B\!+\!n}{l}\left({2^{R_s}\!\!-\!\!1}\right)^l
\Gamma_G(N_A m_B \!+\! N_E\ddot{m}_E \!+\!n\!-\!l)}
{\left(\overline{\gamma}_B \ddot{\overline{\gamma}}_E\right)^{l}\!{2^{l R_s}}\!\left(2^{R_s}m_B \ddot{\overline{\gamma}}_E + \ddot{m}_E \overline{\gamma}_B \right)^{N_A m_B \!+\! N_E\ddot{m}_E \!+\!n\!-\!l}}. \notag
\end{align}

%rrdel We would like to highlight that this is the first time to examine the secrecy performance of the full-CSI scheme for Rician fading channels.
As per \eqref{secercy_outage_MRT}, we know that the secrecy outage probability of the full-CSI scheme is dependent on the locations of both Bob and Eve. This means that we require Bob and Eve's locations for the secrecy performance analysis of the full-CSI scheme. The locations of both Bob and Eve are not only required by the LBB scheme. We note that the secrecy diversity order the full-CSI scheme is also $N_A m_B$ (full diversity order).
Again following a similar procedure of deriving $P_{non}$ in Corollary~2, the probability of non-zero secrecy capacity of the full-CSI scheme is derived as
\begin{align}\label{non-zero_MRT}
\begin{split}
P_{non}^{\mathbf{CSI}} &= 1 - \frac{m_B^{N_A m_B}\ddot{m}_E^{N_E\ddot{m}_E}}{\Gamma(N_E \ddot{m}_E) \ddot{\overline{\gamma}}_E^{-N_A m_B}\overline{\gamma}_B^{-N_E \ddot{m}_E}}\times  \\
&~~~~~~~~~~~~ \sum_{n = 0}^{+\infty}\frac{m_B^{n} \ddot{\overline{\gamma}}_E^n}{\Gamma(N_A m_B + n+1)}\times \\
%===================================================================
&~~~~~~~~~~~~
\frac{ \Gamma\left( N_A m_B+N_E\ddot{m}_E +n\right)}{\left(m_B \ddot{\overline{\gamma}}_E+\ddot{m}_E \overline{\gamma}_B\right)^{N_A m_B +N_E \ddot{m}_E +n}}.
\end{split}
\end{align}

\end{appendices}

\end{document}